# FAFNIR: strategy and risk reduction in accelerator driven neutron sources for fusion materials irradiation data

Elizabeth Surrey[a], Michael Porton[a], Antonio Caballero[a], Tristan Davenne[b], David Findlay[b], Alan Letchford[b], John Thomason[b], James Marrow[c], Steve Roberts[c], Andrei Seryi[d], Brian Connolly[d], Paul Mummery[e], Hywel Owen[e]

*EURATOM/CCFE, Abingdon, OX14 3DB, UK [b]STFC Rutherford Appleton Laboratory, Harwell, OX11 0QX, UK, [c]University of Oxford, Oxford, OX1 3DP, UK, [d] John Adams Institute, University of Oxford, Oxford, OX1 3DP, UK, [d]University of Birmingham, Edgbaston, B15 2TT, UK, [e]University of Manchester, Manchester, M13 9PL, UK*

*Abstract*— The need to populate the fusion materials engineering data base has long been recognized, the IFMIF facility being the present proposed neutron source for this purpose. Re-evaluation of the regulatory approach for the EU proposed DEMO device shows that the specification of the neutron source can be reduced with respect to IFMIF, allowing lower risk technology solutions to be considered. The justification for this approach is presented and a description of a proposed facility, FAFNIR, is presented with more detailed discussion of the accelerator and target designs.

## 1 Introduction

The EU program for a first demonstration fusion power plant, DEMO, envisages engineering design completed by 2030 followed by a 10 year construction program leading to operation in 2040 [1]. Given the scarcity of fusion relevant irradiation data and the need to eliminate unsuitable materials from the design process, the delay in realizing the International Fusion Materials Irradiation Facility (IFMIF) obviously impacts upon the design program for future power plants. This has prompted a study to investigate the feasibility of constructing an intermediate facility based on (near-) available technology, from which the FAcility for Fusion Neutron Irradiation research (FAFNIR) emerged [2]. It is not the intention that this facility replace IFMIF but rather it will provide an intermediate step to IFMIF from existing irradiation facilities, the latter being generally not suited to fusion applications. The purpose of FAFNIR is to generate data to advance the understanding of irradiation effects on materials in a fusion environment and thus enable early elimination of unsuitable candidates and to support an extensive program of modeling. This paper describes in more detail the technical aspects of the program and, in particular, the strategy of target development outlined in [2].

## 2 Fusion Requirements of a Neutron Source

It was shown in [2] that the qualification of candidate materials up to a full lifetime use equivalent to approximately 150dpa - the original intention of IFMIF - is no longer the prime driver of the materials development program in support of DEMO. Assessing the requirements determined by the regulatory, operational and materials database showed that several of the performance requirements of the source may be relaxed. For example, materials degradation phenomena such as irradiation creep, volumetric swelling, and phase instabilities approach saturation at damage levels above 10dpa [3] a level that can be achieved in a reasonable time frame with a less intense source of 5dpa/fpy. The onset of material embrittlement due to the transmutation production of helium and hydrogen is more difficult to quantify as this is only manifest through indirect evidence such as changes in tensile properties and the ductile to brittle transition temperature (DBTT). Experimental evidence suggests that helium concentrations of 400appm at 15dpa have no effect on the tensile properties at temperatures from 250 to $400^0$C whilst the same concentration would increase the DBTT by some $200^0$C as measured by a Charpy test [4]. Despite this apparent problem to detect embrittlement below 15dpa, valuable early elimination of unsuitable materials can be achieved to minimize risk to the engineering design and development programme. Furthermore, recent EU re-assessment of the damage of DEMO, limiting its availability to 30% [1] relaxed the operating characteristics for the first wall to a probable damage level of 20dpa (steels) over its 3 year in-vessel life. Finally, recent advances in the use of millimeter scale mechanical test specimens [5] allow the high flux irradiation volume to be reduced without loss of statistical value in test results.

Considering these points and the need for data that is relevant to and informs the engineering design phase the purpose of the 14MeV neutron source, by priority, is modified to: (i) demonstrate lifetime integrity of the confinement boundary (the vacuum vessel) materials under relevant neutron spectrum and exposure, (ii) identify new phenomena associated with 14MeV neutron irradiation that may impact on the safety case, necessitating further investigation, (iii) provide significant contributions to the population of the engineering materials database and eliminate unsuitable candidate materials, (iv) validate and calibrate fission and ion irradiation techniques and advance the materials modeling

Author's email: elizabeth.surrey@ccfe.ac.uk

capability for fusion without compromising the validity of the neutron spectrum and (v) provide assurance for protection of investment for stakeholders through development of design codes

When combined, these changes to the specification and role of the neutron source bring realization within the capability of existing, or near-term, accelerator and target technology as proposed for FAFNIR, promoting a change in strategy for the realization of a fusion neutron source.

## 3 FAFNIR

The choice of beam and target elements was made after an exhaustive review of neutron sources including spallation, fission, beam-gas and stripping reactions to select the optimum balance between neutron spectrum and technological maturity [2]. Providing irradiation data on a timescale relevant to DEMO precludes a significant R&D program for the neutron source, so risk reduction was a major consideration in selection of the technology.

The proposed facility, FAFNIR, is based on the C(d,n) stripping reaction of a CW 40MeV $D^+$ beam incident on a graphite target. The technical specification is summarized and compared to IFMIF in Table 1. The reaction and beam energy ensure a neutron spectrum peaked at 14MeV whilst providing improved yield over proton beams of the same current. A phased approach to source intensity is proposed to maintain the beam power on the target within present and near-term capabilities for the short term program. This would allow some irradiation data to be obtained whilst, in parallel, the target development program prepared for an increase in beam power. To achieve this, the accelerator must be designed to operate at 30mA, so that only the target would require development or replacement. If the target development program were to begin at the conceptual design stage, it may be feasible to enter operation directly at the near-term option.

Table 1.Proposed Phases of FAFNIR

|  | IFMIF | FAFNIR | | |
|---|---|---|---|---|
|  |  | **Target Technology Level** | | |
|  |  | *Existing (Default)* | *Near-term (Baseline)* | *Prospective (Upgrade)* |
| *Beam* | 40MeV, 250mA | 40MeV, 2.5mA | 40MeV, 5mA | 40MeV, 30mA |
| *Target* | 10MW | 100kW | 200kW | 1.2MW |
|  | Liquid Li | Solid C rotating | Solid C rotating | Solid C rotating |
|  |  | Single slice | Single/ multi- slice | Various options |
| *Typical dpa/fpy: volume (cm³)* | ≥1: 6000 ≥20: 500 ≥50: 100 | ≥0.6: 100 ≥1: 50 ≥3.8: 10 | ≥1: 150 ≥1.5: 100 ≥4: 25 | ≥5:150 ≥7: 100 ~20: 25 |

### 3.1 FAFNIR Accelerator

The FAFNIR proposed accelerator is a $D^+$ ion source with DC injector of ~90keV beam energy, an RFQ accelerating to ~3MeV and a linear accelerator continuing to 40MeV. At the proposed maximum current of ~30mA CW the beam power would be ~1.2MW. This is substantial compared to other accelerator applications; similar powers have been reached by the Los Alamos (normal-conducting) and SNS (superconducting) linacs and by the PSI cyclotron, but at a much higher energy (and hence at a lower current).

It is expected, from experience at the ISIS accelerator, that an availability (including maintenance periods) of 70% will be achievable after operational shakedown. To avoid prohibitively expensive remote maintenance it is essential to minimize induced radioactivity from beam losses. This hands-on maintenance requirement puts strict limits on allowable beam losses and if the absolute beam loss is to be kept the same as in ISIS, the fractional beam loss in the FAFNIR linac would need to be over 100 times lower. It is possible that beam losses may need to be even lower than this given that the range of nuclear reactions possible for particles lost from the beam in typical accelerator structure materials is wider for deuterons than it is for protons. This requires a well-controlled and low-emittance beam in a generous aperture.

The obvious technology choice for a highly reliable $D^+$ source is the off-resonance ECR source and the leading source of this type is the SILHI source developed by CEA, which has delivered a CW $D^+$ ion beam of 140mA for IFMIF-EVEDA [6].

There are two approaches to designing a CW RFQ: the first is to use a pulsed RFQ design but to adopt a high-capacity water-cooling system giving much greater RF power dissipation, typically 120–130kW/m, as in IPHI [7]. This is challenging to achieve whilst simultaneously maintaining the required structural stability. The second approach is to reduce the power dissipation to ~50–60kW/m, as in PXIE [8], with a consequent reduction in accelerating gradient of ~30%. This results in a longer but more conservative RFQ design. In principle the accelerating gradient can be reduced as much as required, the heat dissipated per unit length varying as the square of the accelerating gradient.

On the basis of clear experience on many existing machines around the world, it is possible to match beam currents up to ~30mA into the subsequent accelerator at a relatively low energy of ~2–3 MeV while controlling the effects of space charge.

The challenges for the linac arise from the requirement for CW operation, controlling the beam losses and the powers dissipated in the structure. In practice, the linac RF frequency will likely be dictated by the availability of suitable CW RF sources, the frequency range of 100–200MHz providing the most options. Conveniently this frequency choice leads to larger structures, which is compatible with the requirement for low beam losses.

CW operation is a strong driver towards superconducting technologies where the savings in installed RF and operating costs can be considerable. Superconducting accelerating structures also typically have larger apertures which help to achieve low beam losses. For ~3–40MeV energies suitable superconducting

Author's email: elizabeth.surrey@ccfe.ac.uk

structures are limited to half-wave and spoke resonators, practical experience of which is limited. The low Technology Readiness Level and considerable overhead in cryo-systems may preclude the use of superconducting technology.

A room-temperature drift tube linac is the most conservative choice although it has a significantly reduced aperture compared with superconducting options. As with the RFQ, CW operation leads to challenges in maintaining thermal stability but again reducing the accelerating gradient to 1.5MV/m with 50kWm⁻¹ heat dissipation mitigates this problem. A 10-m long cavity would give an energy gain of ~15MeV with ~500kW of structure power and ~500kW of beam power.

The high energy beam transport (HEBT) for the 40MeV D⁺ beam will have a FODO focusing structure and techniques for generating a uniform (<10%) beam on target with a square profile ~60×60mm² were discussed in [2]. A choice of focusing elements is required: cold quadrupoles or solenoids allow higher average accelerating gradients to be achieved but introduce extra engineering complexity; warm focusing elements require more cryo-modules and a consequent reduction in the average accelerating gradient. Permanent magnet quadrupoles offer reduced complexity but also reduced flexibility for beam tuning; electromagnetic quadrupoles are more complex and add to the drift tube heat load particularly at low energies but offer significant operational benefits when trying to achieve very low fractional beam losses.

### 3.2 FAFNIR Target

The target is the area of greatest concern and the only area where a prototype program is envisaged. A 40MeV D⁺ CW beam incident on a carbon target deposits its entire energy within 0.5cm of the target surface, as shown in Fig 1. The peak deposition corresponds to 35GWm⁻³ for a 5mA beam current and 210GWm⁻³ for 30mA uniformly distributed in a flat-top circular beam of 30mm radius. Dissipating the resultant high power densities in a stationary target would be extremely difficult. By contrast, a rotating target significantly reduces the average power density (and hence deuteron flux) and provides increased surface area for heat transfer.

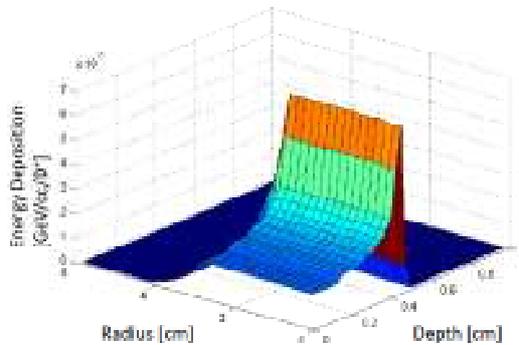

Fig. 1. Deposition profile for 30mA, 40MeV deuteron beam in graphite carbon (FLUKA simulation).

Fig 2 shows the reduction in power density as the radial position, R, of the beam on the target; the corresponding deuteron flux at a radius R=30cm for the

Author's email: elizabeth.surrey@ccfe.ac.uk

5mA beam is 3.2x10¹³cm². The damage to the target depends on the flux and calculations using TRIM suggest that these parameters would produce ~2dpa in the carbon target per six months of operation, implying that target replacement frequency can be combined with other regular maintenance schedules. Clearly, the 30mA beam will require a larger diameter target wheel to avoid frequent replacement of the target.

The proposed target wheel has a metallic hub onto which graphite slices are fitted. The slices are enclosed in a water cooled jacket for most of their circumference, which presents a constant temperature heat sink at ~300K. Assuming slices of thickness 1mm, the majority of the power would be deposited in the fifth slice. A radiation heat transfer model has been used to calculate the time averaged temperature within the annular locus of the beam on the rotating target as a function of beam current and radial position of the beam. The target is assumed to be made of IG43 graphite for which the mechanical properties are well known, although the thermal conductivity was halved to accommodate the anticipated irradiation degradation. The emissivity of the graphite target and the cooling jacket was assumed to be 0.75. The predicted average temperature is 1705K for the 5mA beam incident at a radius of 0.25m and 2056K for the 30mA beam incident at 0.75m. This ignores the effect of temperature pulsing (discussed below) due to the periodic exposure of the rotating graphite to the beam.

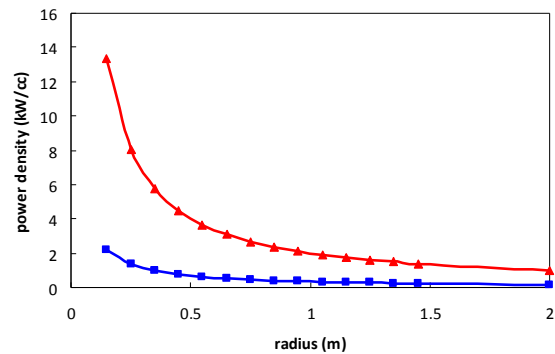

Fig. 2. Power density reduction with increasing radius of impingement. Beam current 5mA ■ , beam current 30mA ▲

The non-uniform heating, both time averaged and transient, induces stress in the annulus as does the rotation which induces tensile stress in the tangential and radial directions. The tangential stress is higher than the radial stress by two orders of magnitude for R=0.75m and given that the tensile stress limit of IG43 graphite is 37MPa becomes a limiting factor for the larger radii considered at high speed. However, irradiation does increase graphite strength and elastic modulus although this has not been considered in this analysis

The combined effects of the thermal and rotational stresses and the average temperature are shown in Fig 3 for 5mA and 30mA beam currents and a rotational speed of 1000rpm. The temperature limit (2000K) due to sublimation and thermal creep and the tensile stress limit (37MPa) are also shown as the broken and dashed curves respectively. The combination of stresses gives an optimum value of radius R for each beam current. For 30mA operation R=0.5m for which the average temperature is 2257K, above the temperature limit for sublimation and thermal creep. Further modelling has shown that introducing a forced flow of helium between the stator and rotor slices can reduce the average temperature to ~1950K whilst also reducing the stress induced by the thermal gradient by 8MPa. This improves the stress design margin from ~80% to ~55% of the tensile limit, albeit at the introduction of additional engineering complexity. Designing the discs to be in compression (the compressive strength of IG43 is 90MPa) may be an alternative solution.

The high operating temperatures are a concern for graphite sublimation and thermal creep rate. Operating at a temperature below 2000K reduces these effects and typical measured values are: for evaporation $<10^{-2}$ mm/month [9] and for creep $<10^{-7} s^{-1}$ at temperatures around 2000K and stress of several MPa [10]. Switching to conical target geometry or replacing the final 1mm disc with multiple, thinner components to distribute the absorbed power are both possibilities. Note that the risk from the reducing thermal conductivity as the carbon becomes radiation-damaged would not be too serious a problem due to the typically short paths for heat conduction and removal. Also, operating graphite at high temperature is thought to have a positive effect in terms of mitigating radiation damage, although the calculations presented here used a thermal conductivity reduced by 50% to account for these effects.

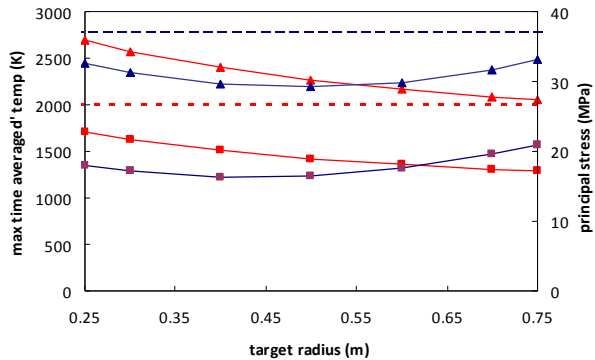

Fig 3 Average temperature and principal stress resulting from thermal and rotational effects as functions of target radius. Beam current 5mA: ■ resultant stress and ■ average temperature. Beam current 30mA ▲ resultant stress and ▲ average temperature. Temperature limit due to sublimation and creep — — and tensile strength limit — — —

The rotation of the target induces an additional temperature effects in the graphite. On each revolution there is an increase in temperature as the graphite passes through the beam spot; during the rest of the revolution that heated point on the annulus will be cooling and the heat is spread through the annular slice. The discrete increase in temperature and the rate of heating depend on the speed of rotation of the wheel as shown in Fig 4. At a few rpm the temperature jump per revolution is of the order 100 to 1000K but decreases rapidly with rotation speed to ~60K at 1000rpm for the 30mA beam with a 0.5m target radius.

Author's email: elizabeth.surrey@ccfe.ac.uk

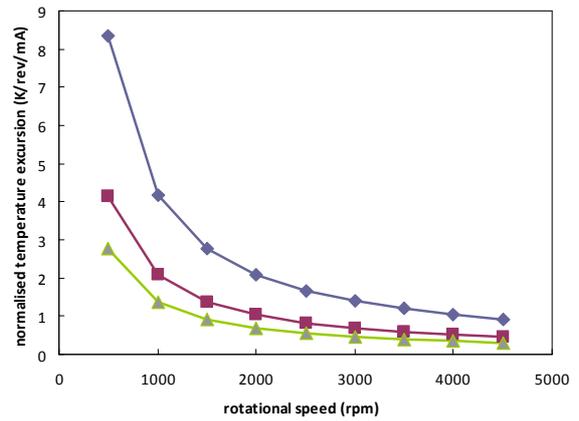

Fig. 4. Peak temperature excursion, normalized to beam current, due to rotation of the target as a function of rotational speed for R=0.25m ♦, R=0.5m ■ and R=0.75m ▲.

Increasing rotation speed reduces the pulsed temperature rise however stress levels are significantly increased at higher speeds. Segmenting the annular slice into sections maybe desirable from a stress point of view but ultimately an optimisation process will be needed to find the best compromise. The rapid heating rates are not high enough to cause significant concern with regard to inertial stresses in the graphite; for inertial stress to be important the heating time needs to be shorter than the expansion time of the target. The timescales for radial and axial expansion of the graphite slice are certainly short with respect to the heating time of ~1ms at 1000rpm. The circumferential expansion time maybe comparable to the heating rates at say 6000rpm which could result in some inertial waves travelling around the circumference of the wheel. However this is not a major concern as segmenting the slice would solve the issue.

The strain induced by the effect of neutron irradiation with a fission spectrum is well known, being a few percent for doses of $\sim 10^{21}$ neutrons $cm^{-2}$. This is comparable for the 30mA beam and the assumed target lifetime of six months but is also temperature dependent. Of course, unknown phenomena associated with 14MeV neutrons may affect the target performance.

Comparing the FAFNIR proposal with existing and near-term target developments, summarized in Table 2, single-slice rotating carbon targets have been successfully and widely deployed with deposited powers up to ~100kW. Therefore the FAFNIR baseline, which seeks to apply existing technology, foresees a 100kW target to accommodate a 40MeV 2.5mA D$^+$ beam. From the results of Fig 3 it appears feasible to consider designing the FAFNIR target for initial operation at 5mA. This would improve the effectiveness of the facility as an irradiation source. In the near term (i.e. within five years), parallel prototyping activities for SPIRAL2 [11] at GANIL (France) and for the FRIB [12] at MSU (USA) will prospectively provide targets capable of handling 200kW of deposited beam power. These programs would be of direct relevance to the FAFNIR target design. Note that these targets will be operating at temperatures higher than shown in Fig 3 for the 30mA FAFNIR option.

Beyond 200kW, a dedicated target program would be required and could be undertaken in parallel to FAFNIR operation with the intention of upgrading the target power handling still further and achieving dpa rates compatible with end-of-life DEMO irradiations. Such irradiations necessitate a 40MeV 30mA D$^+$ beam and ~1.2MW target power. Several target approaches should be assessed but one could perhaps envisage integration of the RIKEN and MSU experiences to achieve an appropriate actively-cooled multi-slice target. Similarly, fluidized powder targets should be explored. Therefore FAFNIR would seek to apply the considerable worldwide experience of operating rotating solid targets for accelerator-based neutron and radio-isotope beam systems to deliver a reliable target. By seeking to collaborate with the various ongoing development programs around the world, the target power handling could then be readily enhanced, both in the near-term and long-term to achieve stepped upgrades in capability and enhanced irradiation conditions within the facility's test volume. It is recognized that existing solid target technology is not adequate for the full performance specification of the facility and that development in this area would be needed. An aggressive target program is therefore proposed that would seek to cooperate with other groups in order to realize a 200kW target and 5mA beam current from the outset of FAFNIR operations. Continuation of that program in parallel to the FAFNIR build, commissioning and early operational activities would enable further rapid upgrade of the facility to achieve the full envisaged capability of 30mA beam current, facilitating irradiations of ~100cm$^3$ of material in excess of 20dpa by 2030.

Table 2.  Existing and near-term target properties

| Facility | Deposited Power (kW/ GWm$^{-3}$) | Temp (K) | Comments |
|---|---|---|---|
| *Existing* | | | |
| LAMPF | 16/ 25 | 1223 | Single-slice, radiation cooled 28rpm radius 0.15m |
| PSI Target E | 60 | 1700 | 60rpm, 1dpa/op yr radius 0.225m |
| RIKEN BigRIPS prototype | 1.8/ 800 (tested) | 570 | Water cooled single slice 100rpm |
| *Near Term* | | | |
| FRIB | 400/ 6x10$^4$ | 2170 | Multi-sliced, radiation cooled, 5000rpm radius 0.15m |
| SPIRAL2 | 200 | 2196 | Single sliced, radiation cooled 50dpa 3000rpm radius 0.06m |
| JPARC | 23.4 | 1010 | Static, He cooled 0.2dpa/yr radius 0.013m |

### 3.3 FAFNIR Irradiation Volume

The indicative dose distribution in the irradiation volume and the strategy for material sample irradiation are described in detail in [2] and are summarized here. The neutron yield was calculated using MCNPX and both the ISABEL model and the TENDL library for the deuteron-nucleon interaction. The results were compared to the experimental measurements of [13] for a 15mm thick carbon target. The ISABEL model was found to give better agreement with the experimental data (within a factor of two over the neutron energy range 5-35MeV) than the TENDL library (a factor of ten difference over the same neutron energy range).

The neutron yield was then computed for the FAFNIR 5mm thick carbon target using MCNPX and ISABEL and the dpa/fpy estimates obtained from the damage energy cross section for $^{56}$Fe in MCNPX, following the NRT model [14]. It should be noted that there is considerable error of approximately 30% in this calculation, arising from uncertainties in the cross section, which together with the error in the neutron yield (largely underestimated by the computation) yields significant error. This is an area for further development in the future, noting that a significant effort was expended on similar calculations for IFMIF e.g.[15].

For the upgraded FAFNIR specification the anticipated dose volume is between 40x40x60mm$^3$ and 50x50x60mm$^3$. Within this volume neutron dose and dose rate will vary greatly with depth; irradiation temperature will also vary to some extent with position but it is expected that this variation can be minimized by use of differential external heating and with careful choice of location of test specimens. Smaller specimen types will be used than those proposed for IFMIF, where possible and where valid data can be obtained [2]. This is a very rapidly developing area of study via experiment and modeling and would be exploited should the issues of transferring data from very small test scales to the engineering scale be resolved. Typical sample sizes could be order of millimeters and use will be made of the "dead zones" in mechanical test samples to provide samples for (nano) indentation testing, the measurement of thermal conductivity, electrical conductivity and swelling, micromechanical test element production capable of providing accurate characterization of additional mechanical properties such as Young's modulus [16] and lift-out samples for transmission electron microscopy (TEM) and atom-probe tomography (APT) specimen preparation.

### 3.4 FAFNIR Infrastructure

A facility such as FAFNIR will need to provide adequate service and analytical capability. To cater for the sample testing substantial active post-irradiation examination facilities will be needed, such as specimen sectioning and polishing, mechanical testing, and focused-ion-beam instruments. Other activities such as APT, TEM and micromechanical testing could be carried out off-site (for example at university departments), since specimen

Author's email: elizabeth.surrey@ccfe.ac.uk

sizes would be small enough to give acceptably low activity per specimen.

In addition, areas for assembly, commissioning and testing of components will be required, particularly for the accelerator RF generators, for which soak testing will be a mitigation strategy to achieve the 70% availability target. Access to high-quality manufacturing capabilities - in-house, locally, nationally, and internationally, and combinations thereof - will be necessary for the same reason. Waste and discharge management, radiolysis of cooling water for corrosion monitoring and production of tritium within the water must be considered along with monitoring, containment and exhaust systems These activities will be continuous throughout the life of the facility.

It is self-evident that it is important to design to minimize running costs, energy consumption and the size of the personnel complement, and to maximize equipment reliability. Of particular operational importance are the choices made for the target station layout, as this can have a very significant effect on the operability of the facility in practice. Involved in the choices are such considerations such as: remote access for active component change, hot cell requirements, design and management of operational activities, cooling plant layout and design (redundant systems, access and control protocols), diagnostic strategy for target systems, and lifting arrangements both in normal and abnormal conditions.

### 3.5 Project Risk, Costs and Timescale

A risk register has been compiled for the major components, including the materials testing program [2]. Unsurprisingly, the most severe risk is associated with the target, viz. structural failure caused by thermal or rotational stress. In mitigation the beam size could be increased or the beam energy could be spread, however both would compromise the neutron flux. This emphasizes the need for a focused design and qualification effort for the target.

The project costs were obtained with reference to the reasonably detailed project breakdown structure for IFMIF given in the CDR [17] and the methodology described in [2], assuming no prior existence of facilities. The costs of providing post-irradiation examination facilities and remote handling hot cells for the targets have been included. The total construction cost in 2012 (including manpower and contingency) was estimated as US$344M for FAFNIR compared to US$1145M for IFMIF. The total commissioning costs of FAFNIR are estimated as US$41M and operational costs are US$12M per year assuming 70% availability, 24 hour operation in parallel with a target program and based on operational costs of ISIS. Construction of the facility, estimated to be of seven years duration, would have to start soon if material irradiation to over 10dpa is to be achieved before commencement of the DEMO construction in 2030.

## 4 Conclusions

The technical specification of the FAFNIR facility has been considered in more detail than in [2] wherein the

Author's email: elizabeth.surrey@ccfe.ac.uk

case for its construction was argued. The issues for the target at highest beam power, identified in [2], have been examined. Operation at 30mA appears challenging in terms of temperature and stress for a carbon-based target, although the analysis presented here is conservative. Several means of mitigating these issues are currently under assessment.

To meet the requirements of the EU Fusion Roadmap [1] and to provide irradiation exposure of tens of dpa, construction will need to begin in a timely manner. Estimates for the cost of FAFNIR are approximately 30% those of IFMIF. FAFNIR would give a lower intensity than IFMIF but is also of lower technological risk, as it exploits existing or near term technologies to avoid lengthy R&D programs and challenging specifications.

FAFNIR, rather than replacing IFMIF, will generate data to support the engineering design phase of DEMO, identify unknown phenomena associated with 14MeV neutron irradiation and support an extensive program of modeling to advance the understanding of irradiation effects on materials in a fusion environment and thus will enable early elimination of unsuitable candidates.

## Acknowledgment

This work was part-funded by the RCUK Energy Program under grant EP/I501045. To obtain further information on the data and models underlying this paper please contact PublicationsManager@ccfe.ac.uk

## References

[1] F Romanelli, Fusion Electricity, EFDA, 2012, ISBN 9783000407208

[2] E Surrey, M Porton, et al., Application of Accelerator Based Neutron Sources in Fusion Materials Research, Proc 23rd Symp on Fusion Engineering, San Francisco, USA, June 2013

[3] E. Gaganidze and Jarir Aktaa, Fus. Eng. Des., **88**, 2013, pp118– 128

[4] S.J. Zinkle and A. Möslang, Fus. Eng. Des **88**, 2013, pp472– 482

[5] D.E.J. Armstrong, A.J. Wilkinson, S.G. Roberts, Materials Research Society Symposium Proceedings "Probing Mechanics at Nanoscale Dimensions, **1185**, 2009, p7-13.

[6] N Chauvin, et al., Beam Commissioning of the linear IFMIF prototype accelerator injector, Proc IPAC2013, Shanghai, China (2013) THPWO006

[7] P.Balleyguier and M.Painchault, Design of RF power input ports for IPHI RFQ, Proc EPAC 2002, Paris, France

[8] S. Virostek, et al, Design and analysis of the PXIE CW RFQ, Proc IPAC 2012, New Orleans, USA (2012) THPPC034

[9] L.B. Tecchio, et al., Measurement of Graphite Evaporation Rate, LNL Annual Report, INFN Laboratori Nazionali di Legnaro, Italy, 2009, p251

[10] P Wagner, A.R. Driesner and L.A. Haskin, J. App. Phys, **30**, 1959, p152

[11] L.B. Tecchio et al. The prototype of the 50kW converter. Spiral2 Conference (2012)

[12] F. Pellemoine et al. Thermo-mechanical behavior of a

single slice test device for the FRIB high power target. Nucl Instr Met Phys Res A **655** (2011) 3-9

[13]  G. Lhersonneau, Nucl. Instrum. and Meth., **A 603**, 2009, pp228-235

[14]  M.J. Norgett, M.T. Robinson and I.M. Torrens, Nucl. Eng. Des., **33**, 1975, pp50.

[15]  U. Fischer, M. Avrigeanu, P. Pereslavtsev, S.P. Simakov, I. Schmuck, J. Nucl. Mat., **367-370**, 2007, pp1531‑1536

[16]  D.E.J. Armstrong, A.J. Wilkinson and S.G. Roberts, Measuring anisotropy in Young's modulus using micromechanical testing in single crystal, J Mater Res, **24**, (2009) 3268-76

[17]  IFMIF International Team, IFMIF Comprehensive Design Report, IEA, 2004

Author's email: elizabeth.surrey@ccfe.ac.uk